\documentclass[final]{svjour3}                     
\smartqed  
\usepackage{graphicx}
\usepackage{amsmath,amsfonts,amssymb}
\usepackage{comment}
\journalname{Journal of Low Temperature Physics - QFS2009}

\begin{document}

\title{Exact Results for Tunneling Problems of Bogoliubov Excitations in the Critical Supercurrent State
}


\author{Daisuke Takahashi$^1$          \and
        Yusuke Kato$^1$ 
}


\institute{1:Department of Basic Science, University of Tokyo, \at
              Tokyo, 153-8902, Japan. \\
              \email{takahashi@vortex.c.u-tokyo.ac.jp}           
}

\date{Received: date / Accepted: date}

\maketitle

\begin{abstract}
	We show the exact solution of Bogoliubov equations at zero-energy in the critical supercurrent state for arbitrary shape of potential barrier. With use of this solution, we prove the absence of perfect transmission of excitations in the low-energy limit by giving the explicit expression of transmission coefficient. The origin of disappearance of perfect transmission is the emergence of zero-energy density fluctuation near the potential barrier.
\end{abstract}

\section{Introduction}\label{sec:intro}
	In 2001-2003, Kovrizhin and his collaborators\cite{kov} have shown an interesting theoretical prediction on the tunneling properties of Bogoliubov excitations\cite{Bogoliubov}; the Bogoliubov excitations show perfect transmission across a potential barrier in the low-energy limit. It is called anomalous tunneling. It is quite different from an ordinary particle obeying Schr\"odinger equation, which shows perfect reflection in the low-energy limit. Later, Danshita \textit{et al.}\cite{danshita2} have extended the problem in the presence of the supercurrent. Solving the delta-functional barrier problem, they have shown that (a) perfect transmission occurs even when the condensate supercurrent exists, except for the critical supercurrent state; (b) under the critical supercurrent, the partial transmission occurs. Thus, a consistent explanation for both perfect transmission in the non-critical states and the absence of perfect transmission in the critical state had been highly desired. As for the perfect transmission in the non-critical states, its physical mechanism has been investigated in many works\cite{kato,ohashi} and importance of the similarity between the condensate and low-energy excitations has been pointed out. However, few results have been known on the critical supercurrent state. It is crucial to understand the reason why the critical supercurrent state shows an exceptional behavior, for the purpose of constructing a unified picture for non-critical and critical states. In our study, we prove the absence of perfect transmission in the critical supercurrent state for arbitrary shape of barrier, and clarify its physical mechanism.
\section{Formulation}\label{sec:funda}
\begin{figure}[tb]
	\begin{minipage}{0.58\textwidth}
		\includegraphics[scale=0.75]{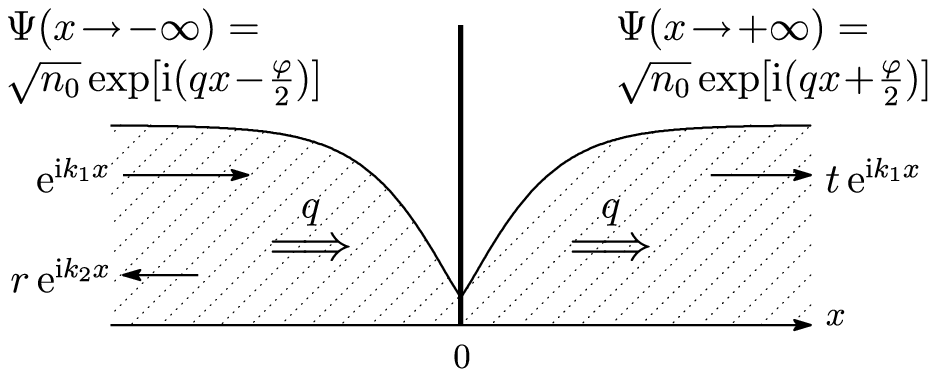}
	\end{minipage}
	\hfill
	\begin{minipage}{0.39\textwidth}
		\includegraphics[scale=1.1]{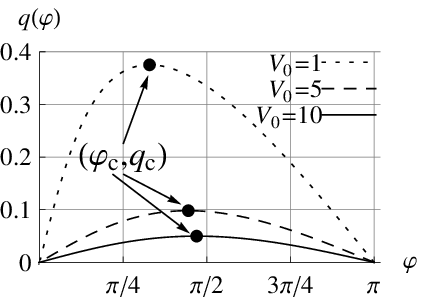}
	\end{minipage}
	\begin{minipage}[t]{0.55\textwidth}
		\caption{\label{fig:curr}Supercurrent state of the condensate wavefunction and tunneling problem of Bogoliubov excitations across a short-ranged potential barrier. The shaded area shows the condensate density profile $ |\Psi(x)|^2 $, and  $ q $ represents the magnitude of the supercurrent (\textit{not} a wavenumber of a Bogoliubov excitation). Such a reflectionless supercurrent solution must have the phase difference  $ \varphi $, and supercurrent $ q $ depends on $ \varphi $ as shown in Fig.~\ref{fig:cur:joseph2}. In this supercurrent state, we further consider the tunneling problem of Bogoliubov excitations, i.e., find the transmission and reflection amplitudes $ t $ and $ r $ for an incident wave $ \mathrm{e}^{\mathrm{i}k_1x} $.}
	\end{minipage}
	\hfill
	\begin{minipage}[t]{0.41\textwidth}
		\caption{\label{fig:cur:joseph2}Josephson relation  $ q(\varphi) $ for a delta functional barrier $ U(x)=V_0\delta(x) $ with $ V_0=1,\,5,\,\text{and }10 $. Though the exact expression of $ q(\varphi) $ is complicated, when $ V_0\gg 1 $, it approximately becomes $ q(\varphi)\simeq \frac{1}{2V_0}\sin\varphi+O(\frac{1}{V_0{}^2}) $\cite{danshita2}. $ (\varphi_{\text{c}},q_{\text{c}}) $ are the critical supercurrent states, where the supercurrent reaches to the maximum value. An example of Josephson relation for a rectangular barrier can be seen in Ref.~\cite{baratoff}.}
	\end{minipage}
	\end{figure}

	We begin with time-dependent Gross-Pitaevskii(GP) equation
	\begin{align}
	\mathrm{i}\frac{\partial}{\partial t}\psi(x,t)=\left(-\frac{1}{2}\frac{\partial^2}{\partial x^2}+U(x)\right)\psi(x,t)+|\psi(x,t)|^2\psi(x,t).
	\end{align}
	Here we use a dimensionless description. Assuming the solution in the form of
	\begin{align}
	\psi(x,t)={\rm e}^{-\mathrm{i} \mu t}\!\left\{\Psi(x)+\left[u(x)\mathrm{e}^{-\mathrm{i} \epsilon t}-v^*(x)\,\mathrm{e}^{\mathrm{i} \epsilon t}\right]\right\}, \label{eq: smallpsi}
	\end{align}
	and ignoring the higher-order terms of $ u $ and $ v $, we obtain stationary GP equation
	\begin{align}
		\hat{L}\Psi(x)=0,\ \hat{L} = -\frac{1}{2}\frac{\mathrm{d}^2 }{\mathrm{d} x^2} +U(x)-\mu+ |\Psi(x)|^2 \label{eqGP}
	\end{align}
	for the condensate wavefunction, and Bogoliubov equations
	\begin{gather}
		\begin{pmatrix}\hat{L}+|\Psi(x)|^2 & -(\Psi(x))^2 \\ -(\Psi(x)^*)^2 & \hat{L}+|\Psi(x)|^2  \end{pmatrix} \begin{pmatrix} u(x) \\ v(x) \end{pmatrix} = \epsilon \begin{pmatrix} u(x) \\ -v(x) \end{pmatrix} \label{eqBogo}
	\end{gather}
	for wavefunctions of Bogoliubov excitations.\\
	\indent Henceforth, we would like to consider the problem depicted in Fig.~\ref{fig:curr}. Therefore, we assume that $ U(x) $ is a short-ranged potential barrier, and that the condensate wavefunction has the following asymptotic form:
	\begin{align}
		\Psi(x\rightarrow\pm\infty) = \exp\Big[\mathrm{i}\Big(qx\pm\frac{\varphi}{2}+\text{const.}\Big)\Big]. \label{eqGPasymp}
	\end{align}
	This asymptotic behavior determines chemical potential as $ \mu=1+q^2/2 $. Setting  $ \Psi(x)=A(x)\exp[\mathrm{i}\Theta(x)] $, one obtains
	\begin{gather}
		\hat{H}A=0,\ \hat{H} = -\frac{1}{2}\frac{\mathrm{d}^2 }{\mathrm{d}x^2}+U+\frac{q^2}{2\,}\Bigl(\frac{1}{A^4}-1 \Bigr)-1+A^2, \label{eqGPamp} \\
		A^2\frac{\mathrm{d} \Theta}{\mathrm{d} x} = q.\label{eq: Theta}
	\end{gather}
	Here the second equation is already integrated once, and a constant of integration becomes $ q $ from Eq.~(\ref{eqGPasymp}). Figure~\ref{fig:cur:joseph2} shows the Josephson relation. From this figure, the condition for the critical supercurrent state is given by
	\begin{align}
		\frac{\partial q}{\partial\varphi}=0. \label{cricondition}
	\end{align}
	By means of the condensate phase $ \Theta $, we introduce the following quantities:
	\begin{align}
	S=u\mathrm{e}^{-\mathrm{i}\Theta}\!+v\mathrm{e}^{\mathrm{i}\Theta}, \quad G=u\mathrm{e}^{-\mathrm{i}\Theta}\!-v\mathrm{e}^{\mathrm{i}\Theta}. \label{eq: uv-SG}
	\end{align}
	Bogoliubov equations are then rewritten as
	\begin{align}
		\hat{H}S-\frac{\mathrm{i}q}{A}\frac{\mathrm{d}}{\mathrm{d}x}\!\left( \frac{G}{A} \right)\! = \epsilon G, \label{eq:bogos} \\
		(\hat{H}+2A^2)G-\frac{\mathrm{i}q}{A}\frac{\mathrm{d}}{\mathrm{d}x}\!\left( \frac{S}{A} \right)\! = \epsilon S. \label{eq:bogog}
	\end{align}
	$ S $ and $ G $ can be interpreted as phase and density fluctuations, because one can show the following expressions from Eq.~(\ref{eq: smallpsi}):
	\begin{align}
		|\psi|^2 &= A^2 \Bigl[1+\frac2A \operatorname{Re}(G{\rm e}^{-\mathrm{i}\epsilon t})\Bigr]+O(S,G)^2, \label{eq:densityfl} \\
		\frac{\psi}{|\psi|}&={\rm e}^{-\mathrm{i}\mu t+\mathrm{i}\Theta}\Bigl[1+\frac{\mathrm{i}}{A}\operatorname{Im}(S{\rm e}^{-\mathrm{i}\epsilon t})\Bigr]+O(S,G)^2. \label{eq:phasefl}
	\end{align}
	\indent In order to obtain the transmission amplitude, we construct the tunneling solution with the following asymptotic form:
	\begin{align}
		\begin{pmatrix}S \\ G \end{pmatrix} = \left\{ \begin{aligned}&  \begin{pmatrix} 1 \\ k_1^2/[2(\epsilon-qk_1)] \end{pmatrix}\mathrm{e}^{\mathrm{i}k_1x}+ \begin{pmatrix} 1 \\ k_2^2/[2(\epsilon-qk_2)] \end{pmatrix} \tilde{r}\,\mathrm{e}^{\mathrm{i}k_2x} & (x\rightarrow-\infty)\hphantom{.} \\ & \begin{pmatrix} 1 \\ k_1^2/[2(\epsilon-qk_1)] \end{pmatrix}t\,\mathrm{e}^{\mathrm{i}k_1x} & (x\rightarrow+\infty). \end{aligned} \right. \label{eq:recasym}
	\end{align}
	Here $ k_1 $ and $ k_2 $ are real positive and negative roots of the dispersion relation $ \epsilon=qk+\frac{1}{2}\sqrt{k^2(k^2+4)} $.
\section{Exact Solution of Bogoliubov Equations at $ \epsilon=0 $ in the Critical Supercurrent State for Arbitrary Shape of Potential Barrier}\label{sec:exact}
\begin{figure}[tb]
		\begin{center}
		\begin{tabular}[c]{c}
		\includegraphics[scale=1.0]{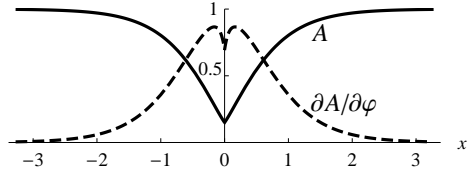}
		\end{tabular}
		\caption{\label{fig:curr:loc} Localized density fluctuation solution $ G=\partial A/\partial\varphi $ for a delta functional barrier $ U(x)=V_0\delta(x) \text{ with } V_0=4.6 $. (We note that $ \partial A/\partial\varphi $ in the figure is multiplied by a constant.)}
		\end{center}
	\end{figure}

	Since Bogoliubov equations are two-component and second-order linear differential equations, the general solution can be expressed in terms of four linearly-independent solutions. We have obtained the exact solutions of Bogoliubov equations at $ \epsilon=0 $ valid only for the critical current state, in other words, only when the condition (\ref{cricondition}) holds. They are given by \cite{TakahashiKato}
	\begin{gather}
	\begin{split}
		\begin{pmatrix} S_{\text{I}}  \\  G_{\text{I}} \end{pmatrix} &= \begin{pmatrix}\, A \, \\ 0 \end{pmatrix}\!, \qquad\qquad \begin{pmatrix} S_{\text{II}} \\ G_{\text{II}} \end{pmatrix} = \begin{pmatrix}\displaystyle \! A\!\int^x_0\!\frac{\mathrm{d}x}{A^2}-2\mathrm{i}qA\!\int^x_0\!\frac{G_{\text{II}}\mathrm{d}x}{A^3} \\[2ex] \displaystyle -2\mathrm{i}qA_\varphi\!\int^x_0\!\frac{A_3\mathrm{d}x}{A_{\varphi}^2} \end{pmatrix}\!, \\
		\begin{pmatrix} S_{\text{III}}  \\  G_{\text{III}} \end{pmatrix} &= \begin{pmatrix}\displaystyle -2\mathrm{i}qAA_3 \\ \displaystyle A_\varphi \end{pmatrix}\!, \!\quad \begin{pmatrix} S_{\text{IV}}  \\  G_{\text{IV}} \end{pmatrix} = \begin{pmatrix}\displaystyle \!-2\mathrm{i}qA\!\int^x_0\!\frac{G_{\text{IV}}\mathrm{d}x}{A^3} \\[2ex] \displaystyle A_\varphi\!\int^x_0\!\frac{\mathrm{d}x}{A_\varphi^2} \end{pmatrix}\!.
	\end{split}\label{crizero}
	\end{gather}
	Here we have introduced the following notations:
	\begin{align}
		A_{\varphi} := \frac{\partial A}{\partial\varphi},\quad A_3(x) := \int_0^x\!\frac{A_{\varphi}(x')\mathrm{d}x'}{A(x')^3}.
	\end{align}
	$ (S_{\text{I}},G_{\text{I}}) $, which can be expressed as $ (u,v)=(\Psi,\Psi^*) $, is a well-known solution\cite{fetter}. The solution $ (S_{\text{III}},G_{\text{III}}) $, which can be expressed as  $ (u,v)=(\partial\Psi/\partial\varphi,-\partial\Psi^*/\partial\varphi) $, is specific to the critical current state, and represents the localized density fluctuation near the potential barrier. See Fig.~\ref{fig:curr:loc}.  $ (S_{\mathrm{II}},G_{\mathrm{II}}) $ and  $ (S_{\mathrm{IV}},G_{\mathrm{IV}}) $ are exponentially divergent solutions. However, as we will see in the next section, \textit{all four solutions are necessary to prove the absence of perfect transmission.}
\section{Absence of Perfect Transmission: Sketch of Proof}
	\indent Using the exact zero-energy solution shown in the previous section, we can prove the absence of perfect transmission. In the following, we show the sketch of proof. The detailed calculation is given in Ref.~\cite{TakahashiKato}. Henceforth, we assume $U(x)=U(-x)$ for simplicity. \\  
	\indent i) \ Our goal is to obtain the tunneling solution (\ref{eq:recasym}) up to first-order in $ \epsilon $, that is, 
		\begin{align}
			S(x) \longrightarrow \begin{cases} 1+\tilde{r}^{(0)}+\epsilon\left( \tilde{r}^{(1)}+\left( \frac{\mathrm{i}}{1+q}+\frac{\hphantom{{}^{(0)}}\mathrm{i}\tilde{r}^{(0)}}{-1+q} \right)x \right)+O(\epsilon^2) & (x\rightarrow-\infty) \\[2ex] t^{(0)}+\epsilon\left( t^{(1)}+\frac{\hphantom{{}^{(0)}}\mathrm{i}t^{(0)}}{1+q}x \right)+O(\epsilon^2) & (x\rightarrow+\infty). \end{cases} \label{eq:asympo}
		\end{align}
		Here $ t=t^{(0)}\!+\!\epsilon t^{(1)}\!+\!\dotsb, $ and $ \tilde{r}=\tilde{r}^{(0)}\!+\!\epsilon \tilde{r}^{(1)}\!+\!\dotsb. $ \\ 
	\indent ii) \ In order to achieve the above purpose, we construct the solution of Bogoliubov equations in the form of power series with respect to $ \epsilon $:  $ (S,G) = \sum_{n=0}^\infty \epsilon^n (S^{(n)}\!,G^{(n)}) $. We then obtain the following inhomogeneous differential equations: 
		\begin{align}
			\hat{H}S^{(n)}-\frac{\mathrm{i}q}{A}\frac{\mathrm{d}}{\mathrm{d}x}\!\biggl( \frac{G^{(n)}}{A} \biggr)\! = G^{(n-1)}, \label{eqbogoS17}\\
			(\hat{H}+2A^2)G^{(n)}-\frac{\mathrm{i}q}{A}\frac{\mathrm{d}}{\mathrm{d}x}\!\biggl( \frac{S^{(n)}}{A} \biggr)\! = S^{(n-1)}, \label{eqbogoG17}
		\end{align}
		where the right hand sides should read as zero if $ n=0 $. Since four homogeneous solutions are already given in Eq.~(\ref{crizero}), a particular solution can be found by the method of variation of parameters. Thus, $ (S^{(n)}\!,G^{(n)}) $ can be determined from $ (S^{(n-1)}\!,G^{(n-1)}) $ recursively. \\
	\indent iii) \ Bogoliubov equations with finite energy $ \epsilon $ have four linearly-independent solutions. Generally, two of the four behave as plane waves far from the potential barrier, and the other two are unphysical solutions which diverge exponentially. In solving the tunneling problem, we are particularly interested in the former two solutions.\\
	\indent iv) \ Therefore, we extend the non-divergent zero-energy solutions, i.e.,  $ (S_{\mathrm{I}},G_{\mathrm{I}}) $ and  $ (S_{\mathrm{III}},G_{\mathrm{III}}) $,  to first-order in $ \epsilon $. The particular solution of Eqs.~(\ref{eqbogoS17}) and (\ref{eqbogoG17})  obtained by the method of variation of parameters diverges exponentially, so we must cancel the divergent term by adding the divergent homogeneous solutions $ (S_{\mathrm{II}},G_{\mathrm{II}}) $ and  $ (S_{\mathrm{IV}},G_{\mathrm{IV}}) $. This manipulation is most important. After the calculation, we can obtain the asymptotic forms of the solutions extended up to first-order as follows:
		\begin{align}
			S^{\text{total}}_{\text{I}}(x) \longrightarrow 1+\epsilon \left( \frac{q^2-\eta}{\mathrm{i}q(1-q^2)}x+\tilde{\gamma}\operatorname{sgn}x \right)+O(\epsilon^2), \label{nondiv1}\\
			\text{const.}\times S^{\text{total}}_{\text{III}}(x) \longrightarrow \operatorname{sgn}x +\epsilon\left( \frac{q^2+\eta}{\mathrm{i}q(1-q^2)}|x|+\tilde{\lambda} \right)+O(\epsilon^2). \label{nondiv2}
		\end{align}
		Here $ \tilde{\gamma} $ and $ \tilde{\lambda} $ are constants, and $ \eta $ is defined by the following integral:
		\begin{align}
			\eta:=\left[\int_0^{\infty}\!\!\mathrm{d}x\,A\frac{\partial A}{\partial \varphi}\right]\bigg/\left[\displaystyle\int_0^{\infty}\!\!\mathrm{d}x\,\frac{1}{A^3}\frac{\partial A}{\partial \varphi}\right].
		\end{align}\\
	\indent v) \ By making a linear combination of $ S^{\text{total}}_{\text{I}} $ and $ S^{\text{total}}_{\text{III}} $, we can construct the solution of the form (\ref{eq:asympo}). The transmission amplitude is obtained explicitly as
		\begin{gather}
			t^{(0)} = t(\epsilon\rightarrow 0) = \frac{2q\eta}{q^2+\eta^2}.
		\end{gather}
		Unless $ \eta\ne \pm q $,  $ 0<|t^{(0)}|^2<1 $ holds. Thus, the absence of perfect transmission is proved. We have confirmed in the delta-functional barrier model that $ \eta=q $ occurs only when there is no potential barrier.
\section{Discussion}\label{sec:dis}
	\indent Let us consider the physical origin of the absence of perfect transmission. The tunneling solution in the low-energy limit can be written as \cite{TakahashiKato}
	\begin{align}
		\lim_{\epsilon\rightarrow 0}\begin{pmatrix} u \\ v \end{pmatrix} \propto \begin{pmatrix} \Psi^{\hphantom{*}}\! \\ \Psi^*\! \end{pmatrix} -2\mathrm{i}\,\frac{q\!-\!\eta}{q\!+\!\eta}\,\frac{\partial }{\partial\varphi}\!\!\begin{pmatrix}\Psi^{\hphantom{*}}\! \\ -\Psi^*\! \end{pmatrix} \neq \begin{pmatrix} \Psi^{\hphantom{*}}\! \\ \Psi^*\! \end{pmatrix}.
	\end{align}
	This is drastically different from the non-critical states, in which the wavefunctions of excitations coincide with the condensate wavefunction\cite{kato,gencar}. In the non-critical states, there is only one non-divergent solution $ (u,v)=(\Psi,\Psi^*) \leftrightarrow (S,G)=(A,0) $ at $ \epsilon=0 $, so $G$ cannot contribute to the wavefunction of excitations in the low-energy limit. 
	 In the critical supercurrent state, on the other hand, another non-divergent solution  $(u,v)=(\partial\Psi/\partial\varphi,-\partial\Psi^*/\partial\varphi) \leftrightarrow (S,G)=(-2\mathrm{i}qAA_3,A_{\varphi}) $ arises, and contributes to the low-energy wavefunction of excitations. As shown in Fig.~\ref{fig:curr:loc}, this solution represents the density fluctuation localized near the potential barrier. Thus, the presence of local density fluctuation near the barrier in the low-energy limit is the origin of the absence of perfect transmission.\\ 
	\indent We further note that the critical supercurrent states are at the ``phase boundary'' which separates steady flow states and nonstationary flow states. (See, e.g., Fig.~1 in~Ref.~\cite{Pavloff}.) Therefore, our result suggests that the emergence of low-energy density fluctuation can characterize the destabilization of the superflow. In Ref.~\cite{Hakim}, it has been shown by numerical simulation of time-dependent GP equation that if $ q\ge q_{\text{c}} $, soliton-phonon creation occurs and stationary superflow no longer exists. We expect that the local density fluctuation $ G=\partial A/\partial\varphi $ possesses the information on the ``manner of collapse of stationary superflow''. That is, the profile of the density fluctuation could describe a part of collapse phenomena, such as the soliton-phonon creation. Thus, the investigation of the role of low-energy density fluctuation near the critical supercurrent state is left as a future work.
\vspace{-0.1em}
\section{Conclusion}
\vspace{-0.2em}
	In conclusion, with use of the exact solution of Bogoliubov equations at zero-energy, we have exactly proved the absence of perfect transmission of Bogoliubov excitations in the critical supercurrent state for arbitrary shape of potential barrier. The origin of the absence of perfect transmission is the emergence of low-energy density fluctuation near the potential barrier. Because of this density fluctuation, wavefunctions of excitations does not coincide with the condensate wavefunction in the low-energy limit.
\vspace{-0.1em}
\begin{acknowledgements}
This research was partially supported by the Ministry of Education, Science, Sports and Culture, Grant-in-Aid for Scientific Research on Priority Areas, 20029007, and also supported by Japan Society of Promotion of Science, Grant-in-Aid for Scientific Research (C), 21540352.
\end{acknowledgements}



\end{document}